\documentclass[letterpaper]{IEEEtran}
\usepackage{graphicx,times, amsmath, amsfonts,comment}
\usepackage{amssymb,epstopdf}
\usepackage[noend]{algorithmic}
\usepackage{algorithm}

\newcommand{\beq}{\begin{equation}}
\newcommand{\eeq}{\end{equation}}

\newcommand{\tit}{\textit}

\newcommand{\ud}{\mathrm{d}}

\newcommand {\Ebb}{\mathbb{E}}

\begin{document}

\title{Integral Approximations for Coverage Probability}
\author{\IEEEauthorblockN{Sudarshan Guruacharya, Hina Tabassum, and Ekram Hossain \\} }
\maketitle

\begin{abstract}
This letter gives approximations to an integral appearing in the formula for downlink coverage probability of a typical user in Poisson point process (PPP) based stochastic geometry frameworks of the form $\int_0^\infty \exp\{ - (Ax + B x^{\alpha/2})  \} \ud x$. Four different approximations are studied. For systems that are interference-limited or noise-limited, conditions are identified when the approximations are valid. For intermediate cases, we recommend the use of Laplace approximation. Numerical results validate the accuracy of the approximations.
\end{abstract}


\section{Introduction}
In \cite[Th. 1]{Andrews2011}, the authors derived the coverage probability for the downlink transmission of a typical user in the single-tier multi-cell network by assuming that the desired channel undergoes Rayleigh fading and that the base stations (BSs) are spatially distributed according to homogeneous PPP as:
\begin{equation}
p_c = \pi \lambda \underbrace{\int_0^\infty \exp\{ - (Ax + B x^{\alpha/2})  \} \ud x}_{\text{$I$}},
\label{eqn:coverage}
\end{equation}
where $A = \pi \lambda \beta $ and $B = \mu T \sigma^2$ are real non-negative quantities\footnote{A similar formula holds for $K$-tier systems as well \cite{Dhillon2012}.}. Here, $\alpha$ is the path-loss exponent, $\lambda$ is the intensity of the PPP, $T$ is the signal-to-interference-plus-noise ratio (SINR) threshold, $1/\mu$ is the constant transmit power, and $\sigma^2$ is the noise variance. The value of $\beta$ is given by the formula: $\beta = \frac{2(\mu T)^{2/\alpha}}{\alpha} \Ebb_g[g^{2/\alpha}(\Gamma(-2/\alpha,\mu T g) - \Gamma(-2/\alpha))],$ 
where $\Gamma(z)$ is Gamma function while $\Gamma(a,z) = \int_z^\infty x^{a-1}e^{-x}\ud x$ is the upper incomplete Gamma function. The $\Ebb_g[\cdot]$ is expectation taken with respect to interferers' channel distribution $g$. 

The integral $I$ does not have closed-form expression for arbitrary values of $\alpha$. For special cases: 
\begin{align*}
I & = \frac{1}{A + B}, & \mbox{for} \quad \alpha = 2 \\
I & = \sqrt{\frac{\pi}{B}} \exp\Big\{\frac{A^2}{4B}\Big\} Q\Big(\frac{A}{\sqrt{2B}}\Big), & \mbox{for} \quad \alpha = 4
\end{align*}
where $Q(\cdot)$ is the Q-function. 

In practice, the value of $\alpha$ can range anywhere from 1.6 to 6.5, depending on the environment. The path-loss exponent for outdoor urban area is around 3.7 to 6.5; while for indoor single floor office buildings, it is around 1.6 to 3.5; for home environment, it is 3; whereas for stores, it can be from 1.8 to 2.2 \cite[p. 47]{Goldsmith2005}. Eq. (\ref{eqn:coverage}) can be directly evaluated by numerical integration techniques, but qualitative insight is lost in this process. This can instead be replaced by approximations, which allows sufficiently accurate quantitative predictions, while at the same time offers qualitative insight into the relationship between various parameters. Thus, the motivation of this letter is to find simple expressions that allow us to approximate the value of coverage probability for these diverse path-loss exponents.  

In this letter, we investigate a few approximations related to the integral (\ref{eqn:coverage}) and discuss their convergence properties. While we focus on cases when $\alpha > 2$, much of the analysis carries over to cases when $\alpha < 2$ as well.


\section{Integral Approximations}
\subsection{Limiting Cases}
There can be two limiting cases for the integral (\ref{eqn:coverage}): when $A = 0$ and when $B=0$. Although the solutions for these two cases are trivial, they have important physical significance. Physically, $A=0$ can occur when the BS intensity $\lambda \rightarrow 0$ and hence the term $Ax$ in the integral can be neglected. This is referred to as the \tit{noise-limited case}. Similarly, $B=0$ can occur when $\sigma^2 \rightarrow 0$, 
so that the term $Bx^{\alpha/2}$ can be neglected. This is referred to as the \tit{interference-limited case}. These cases can also be used as initial approximations when $A<<B$ or $B<<A$.  

For these two cases, the integral (\ref{eqn:coverage}) assumes simple closed form solutions:
\begin{align}
I & = \frac{2}{\alpha B^{2/\alpha}} \Gamma \Big(\frac{2}{\alpha} \Big), & \mbox{for} \quad A = 0,  \label{eqn:limit-A-zero} \\
I &= \frac{1}{A}, & \mbox{for} \quad B = 0  \label{eqn:limit-B-zero}. 
\end{align}   

To evaluate the integral for the case when $A=0$, we have used the fact that  
$\int_0^\infty x^n e^{-ax^b} \ud x = \frac{1}{b} a^{-\frac{n+1}{b}} \Gamma\Big(\frac{n+1}{b}\Big) $, 
 where $n \geq 0$, $a > 0$, $b > 0$ \cite[Ch. 3.326, Eqn. 2, p. 337]{Jeffrey-Zwillinger2007}. We will use this formula frequently in the rest of the letter. 

The simplicity of these expressions inspires the following simple, closed form of limiting approximation for the cases when both $A$ and $B$ are non-zero:
\begin{equation}
I \approx \Big[A +  \frac{\alpha}{2} \frac{B^{2/\alpha}}{\Gamma(\frac{2}{\alpha})} \Big]^{-1}.
\label{eqn:limiting-approx}
\end{equation}
We see that (\ref{eqn:limiting-approx}) reduces to one of the cases given in (\ref{eqn:limit-A-zero}) or (\ref{eqn:limit-B-zero}) as $A \rightarrow 0$ or $B \rightarrow 0$, respectively, and is exact for $\alpha = 2$. The formula can be used when $A$ and $B$ are comparable to each other, but at the expense of accuracy. 
 

\subsection{Interference-Limited Case (When $B << A$)}
For the case when both $A$ and $B$ are positive, one possible strategy of arriving at an integral approximation is to expand one of the exponential terms as
$ e^{-u} = \sum_{k=0}^n \frac{(-u)^k}{k!} + R_n(u), $
where the Lagrange form of the remainder is 
$ R_n(u) = \frac{(-1)^{n+1} e^{-\xi}}{(n+1)!} u^{n+1},$
such that $0 < \xi < u$. Since $e^{-\xi} \leq 1$, absolute value of  $R_n$ can be upper bounded by
\begin{equation} 
|R_n(u)| \leq \Big|\frac{u^{n+1}}{(n+1)!}\Big| = \frac{|u|^{n+1}}{(n+1)!}. 
\label{eqn:lagrange-bound}
\end{equation}

If we expand the term $\exp(-Bx^{\alpha/2})$ appearing in (\ref{eqn:coverage}) and integrate term wise, we obtain
\begin{align}
I &= \int_0^\infty e^{-Ax} \Big( \sum_{k=0}^n \frac{(-Bx^{\alpha/2})^k}{k!} + R_n(x) \Big) \ud x  \nonumber \\
&= \frac{1}{A} \sum_{k=0}^n \frac{1}{k!} \Big(\frac{-B}{A^{\alpha/2}} \Big)^k \Gamma\Big(\frac{k\alpha}{2} + 1\Big) + I_R, \label{eqn:first-approx}
\end{align}
where the remainder term $I_R = \int_0^\infty e^{-Ax} R_n(x) \ud x$. Neglecting the remainder term $I_R$ gives us our first approximation
\begin{equation}
I \approx \frac{1}{A} \sum_{k=0}^n \frac{1}{k!} \Big(\frac{-B}{A^{\alpha/2}} \Big)^k \Gamma\Big(\frac{k\alpha}{2} + 1\Big).
\label{eqn:first-approx-series}
\end{equation}

Eq. (\ref{eqn:first-approx-series}) reduces to (\ref{eqn:limit-B-zero}) when $n=0$, as such it is a refinement of (\ref{eqn:limit-B-zero}). The authors in \cite{Andrews2011} arrived at the first two terms of the series (\ref{eqn:first-approx-series}) using integration by parts; but they did not elaborate on its validity. If we take infinite number of terms, the series (\ref{eqn:first-approx-series}) is not convergent for $\alpha > 2$ for arbitrary values of $A$ and $B$ (see \textbf{Appendix A}). 

Nevertheless, this form of approximation is appropriate when $B << A$. To quantify the region in which this approximation is valid, we first upper bound the remainder term in (\ref{eqn:first-approx}) as
\begin{align}
|I_R| 
&\leq  \int_0^\infty e^{-Ax} | R_n(x) | \ud x   \nonumber \\ 
&\leq \frac{B^{n+1}}{(n+1)!} \int_0^\infty e^{-Ax} x^{\alpha(n+1)/2} \ud x \nonumber \\
&= \frac{1}{(n+1)!} \frac{1}{A} \Big(\frac{B}{A^{\alpha/2}} \Big)^{n+1} \Gamma\Big(\frac{(n+1)\alpha}{2} + 1\Big).
\end{align}
The second inequality is from (\ref{eqn:lagrange-bound}) where $u = Bx^{\alpha/2}$.

If we take $n$ terms of the approximating series, then for any given error tolerance $\epsilon > 0$, we require that the integral error be $|I_R| \leq \epsilon$. Using the upper bound for $|I_R|$ in this expression, we obtain the bound for $B$ in terms of $A$ as
\begin{equation}
B \leq A^{\alpha/2} (\epsilon K_1 A)^{1/(n+1)}, 
\label{eqn:bound-1}
\end{equation}
where $K_1 = (n+1)!/ \Gamma((n+1)\frac{\alpha}{2} + 1)$. 
Substituting the expressions for $A$ and $B$ in (\ref{eqn:bound-1}), we get
\begin{equation}
\sigma^2 \leq \frac{(\pi \lambda \beta)^{\alpha/2}}{\mu T} ( \epsilon K_1 \pi \lambda \beta)^{1/(n+1)}.
\end{equation}
Thus, we obtain the largest noise variance above which the error of  approximation becomes unacceptable for given number of terms $n$ and error tolerance $\epsilon$. 

However, it is not obvious as to what happens as $n$ is increased to infinity.  For this, we have the limit
\begin{equation*}
\lim_{n \rightarrow \infty} \frac{(n+1)!}{\Gamma((n+1)\frac{\alpha}{2} + 1)} 
  = \left\{ \begin{array}{rcl} 
  		  0, & \mbox{for} & \alpha > 2 \\
  		  1, & \mbox{for} & \alpha = 2 \\
  		  \infty, & \mbox{for} & \alpha < 2.
  		  \end{array} \right.	
\end{equation*}
Hence, for $\alpha > 2$, we have the limit $\lim_{n\rightarrow \infty} K_1^{1/(n+1)} = 1.$ Therefore, we have the following largest value of $\sigma^2$ as
\begin{align}
\sigma^2 & \leq \frac{(\pi \lambda \beta)^{\alpha/2}}{\mu T} \lim_{n\rightarrow \infty} (\epsilon K_1 \pi \lambda \beta)^{1/(n+1)}  
  =  \frac{(\pi \lambda \beta)^{\alpha/2}}{\mu T},
\end{align}
below which the approximation will be valid for any given $\epsilon$.


\subsection{Noise-Limited Case (When $A << B$)}
In the integral (\ref{eqn:coverage}), if we consider expanding the term $\exp(-Ax)$ instead and perform term wise integration, we get
\begin{align}
I &= \int_0^\infty \Big( \sum_{k=0}^n \frac{(-Ax)^k}{k!} + R_n(x) \Big) e^{-B x^{\alpha/2}} \ud x \nonumber \\
 & = \frac{2}{\alpha B^{2/\alpha}} \sum_{k=0}^n \frac{1}{k!} \Big( \frac{-A}{B^{2/\alpha}}\Big)^k \Gamma\Big( \frac{2(k+1)}{\alpha}\Big) + I_R, \label{eqn:second-approx}
\end{align}
where the remainder term $I_R = \int_0^\infty R_n(x) e^{-B x^{\alpha/2}} \ud x$. As before, if we neglect the remainder term $I_R$, we obtain our second approximation as 
\begin{equation}
I \approx \frac{2}{\alpha B^{2/\alpha}} \sum_{k=0}^n \frac{1}{k!} \Big( \frac{-A}{B^{2/\alpha}}\Big)^k \Gamma\Big( \frac{2(k+1)}{\alpha}\Big).
\label{eqn:second-approx-series}
\end{equation}
Eq. (\ref{eqn:second-approx-series}) reduces to (\ref{eqn:limit-A-zero}) when $n=0$, as such it is a refinement of (\ref{eqn:limit-A-zero}).

The remainder term in (\ref{eqn:second-approx}) can be bounded as
\begin{align}
|I_R| 
 & \leq  \int_0^\infty |R_n(x) | e^{-Bx^{\alpha/2}}\ud x  \nonumber \\
 & \leq \frac{A^{n+1}}{(n+1)!} \int_0^\infty x^{n+1} e^{-Bx^{\alpha/2}}\ud x \nonumber \\
 & =\frac{1}{(n+1)!} \frac{2}{\alpha B^{2/\alpha}} \Big(\frac{A}{B^{2/\alpha}}\Big)^{n+1} \Gamma \Big( \frac{2(n+2)}{\alpha} \Big).
\end{align}
The second inequality is from (\ref{eqn:lagrange-bound}) where $u=Ax$.

This form of approximation is suitable for cases when $A<<B$. To find the precise region for which this approximation is valid, consider again an error tolerance of $\epsilon > 0$ and $n$ terms of the approximating series. Since we require that the integral error be bounded by the error tolerance, $|I_R| \leq \epsilon$, this leads us to a bound for $B$ in terms of $A$ as
\begin{equation}
B \geq \Big( \frac{A^{n+1}}{\epsilon K_2} \Big)^{\frac{\alpha}{2(n+2)}},
\label{eqn:bound-2}
\end{equation}
where $K_2 = \alpha(n+1)!/(2\Gamma(2(n+2)/\alpha))$. 
Substituting the expressions for $A$ and $B$ gives us the smallest value of noise variance 
\begin{equation}
\sigma^2 \geq \frac{1}{\mu T} \Big( \frac{(\pi \lambda \beta)^{n+1}}{\epsilon K_2} \Big)^{\frac{\alpha}{2(n+2)}},
\end{equation}
below which the error becomes unacceptably large. 

As in previous case, as $n$ tends to infinity, we have
\begin{equation*}
\lim_{n \rightarrow \infty}\frac{(n+1)!}{\Gamma(\frac{2(n+2)}{\alpha})} 
= \left \{ \begin{array}{rcl} 
		 0, & \mbox{for} & \alpha < 2  \\
		 1, & \mbox{for} & \alpha = 2  \\
		 \infty, & \mbox{for} & \alpha > 2.
		\end{array} \right.
\end{equation*}
Hence for $\alpha > 2$, we have $\lim_{n\rightarrow \infty} K_2^{\alpha/(2(n+2))} = \infty$. Thus, the smallest value of $\sigma^2$ above which the approximation will be valid for any $\epsilon$ is 
\begin{equation*}
\sigma^2  \geq \frac{1}{\mu T} \lim_{n\rightarrow \infty} \Big( \frac{(\pi \lambda \beta)^{n+1}}{\epsilon K_2} \Big)^{\frac{\alpha}{2(n+2)}} = 0.
\end{equation*}
This result implies that the infinite series for (\ref{eqn:second-approx-series}) is convergent when $\alpha > 2$. This is indeed the case (see \textbf{Appendix B}). 


\subsection{Laplace Approximation}
Here, we would like to find an approximation for the case when $A$ and $B$ are comparable to each other, that is, when the system is neither noise limited nor interference limited. We will try to obtain an approximation using Laplace's method. Let $h(x) = Ax + Bx^{\alpha/2}$. The unique global minima of $h(x)$ is at $x=0$, which is where this method is usually applied \cite{Wong2001}. If we take the first order Taylor expansion of $h(x)$ about $x=0$, then it will merely result in the approximation $I \approx 1/A$. Thus, it can be advantageous to consider a point other than the global minima for the Taylor expansion. The Taylor expansion of $h(x)$ about $\hat{x} \in (0,\infty)$ is 
\[ h(\hat{x} + y) = h(\hat{x}) + h'(\hat{x}) y + \frac{h''(\hat{x})}{2} y^2 + R_3(y),\]
where $y = x - \hat{x}$. The remainder term in Lagrange form is
\[ R_3(y) = \frac{h^{(3)}(\xi)}{3!} y^3 = B \binom{\alpha/2}{3} \xi^{\alpha/2 - 3} y^3, \]
where $\xi = \hat{x} + \theta y$ for some $0 < \theta < 1$. Changing the variable in (\ref{eqn:coverage}) from $x$ to $y$, we obtain
\begin{align*} 
I 
 &= \int_{-\hat{x}}^\infty \exp\{-( h(\hat{x}) + h'(\hat{x}) y + \frac{h''(\hat{x})}{2} y^2 + R_3(y)) \} \ud y.
\end{align*}

Let $a = h''(\hat{x})/2$, $b = h'(\hat{x})$, and $c = h(\hat{x})$; completing the square, we get
\[ I = \exp\Big\{\frac{b^2}{4a} - c \Big\} \int_{-\hat{x}}^\infty \exp \Big\{ - a \Big(y+\frac{b}{2a}\Big)^2 \Big\} \exp \{-R_3(y)\} \ud y. \]

Now, let $u = R_3(y)$, and expanding its exponent, we have $e^{-u} = 1 + R_1(u)$, where $R_1(u) = - e^{-\eta} u$ such that $0 < \eta < u$. Its absolute value can be upper bounded as $|R_1(u)| = | - e^{-\eta} u| \leq |u|$. Thus the right integral can be split as 
\[ \int_{-\hat{x}}^\infty \exp \Big\{ - a \Big(y+\frac{b}{2a}\Big)^2 \Big\} \ud y + I_R =\sqrt{\frac{\pi}{a}} \; Q(\hat{y}) + I_R, \]
where $\hat{y} = \sqrt{2a}(-\hat{x}+ b/2a)$ and $Q(\hat{y})$ is the Q-function of $\hat{y}$. The remainder term is $ I_R = \int_{-\hat{x}}^\infty \exp\{ - a (y+\frac{b}{2a})^2 \} R_1(u) \ud y$. If we neglect this remainder term, then we obtain our required approximation as
\begin{equation}
I \approx \sqrt{\frac{\pi}{a}}  \; \exp\Big\{\frac{b^2}{4a} - c \Big\}\; Q(\hat{y}).
\label{eqn:third-approx-laplace}
\end{equation}
The expression is exact when $\alpha = 4$. Unlike the previous two approximations, here the accuracy of approximation in general depends on the value of $\hat{x}$ chosen.

We can bound the remainder term as
\begin{align*}
|I_R| 
& \leq \int_{-\hat{x}}^\infty \exp \Big\{ - a \Big(y+\frac{b}{2a}\Big)^2 \Big\} | R_1(u) | \; \ud y  \\
& \leq \int_{-\hat{x}}^\infty \exp \Big\{ - a \Big(y+\frac{b}{2a}\Big)^2 \Big\} |u|\; \ud y \\
& \leq 2 \int_0^\infty \exp \Big\{ - a \Big(y-\frac{b}{2a}\Big)^2 \Big\} |u|\; \ud y. 
\end{align*}
The last step follows as such: since the function to be integrated is non-negative, $\int_{-\hat{x}}^\infty \leq \int_{-\infty}^\infty$.  When $\alpha > 2$ and $\hat{x} > 0$, both $a > 0$ and $b > 0$; hence $b/2a > 0$, and therefore, the maxima of $e^{ - a(y+\frac{b}{2a})^2}$ occurs in negative axis. So $\int_{-\infty}^\infty \leq 2 \int_{-\infty}^0 = 2 \int_0^\infty$, where the last equality follows by changing the variable from $y$ to $y'=-y$. 

We have $ |u| =  B \, \Big| \binom{\alpha/2}{3} \Big| |\xi|^{\alpha/2 - 3} |y|^3 $.
When $\alpha < 6$, $|\xi|^{\alpha/2 - 3}$ is monotonically decreasing and its maximum value is attained at $\xi = \hat{x}$. Putting $K_3 =  B \Big| \binom{\alpha/2}{3} \Big| |\hat{x}|^{\alpha/2 - 3}$, we have 
\[ |I_R| \leq 2 K_3 \int_0^\infty \exp \Big\{ - a \Big(y - \frac{b}{2a}\Big)^2 \Big\} y^3 \; \ud y. \]

Putting $z = y - b/2a$, we obtain
\begin{align*}
|I_R| &\leq 2 K_3 \int_{-b/2a}^\infty e^{-az^2} (z + \frac{b}{2a})^3 \; \ud z  \\
 &= 2 K_3 \sum_{k=0}^3 \binom{3}{k} \Big(\frac{b}{2a}\Big)^k \int_{-b/2a}^\infty e^{-az^2} z^{3-k} \; \ud z  \\
 &= \frac{K_3}{a^2} \sum_{k=0}^3 \binom{3}{k} \Big(\frac{b^2}{4a}\Big)^{k/2} G\Big(k,\frac{b}{2a}\Big),
\end{align*}
where $ G(k,b/2a) = \Gamma ( 2 - \frac{k}{2}) + (-1)^{3-k} \gamma(2 - \frac{k}{2}, \frac{b}{2a})$ and $\gamma(a,z) = \int_0^z x^{a-1}e^{-x}\ud x$ is the lower incomplete gamma function.

Empirical considerations have led us to choose $\hat{x} = (A + B^{2/\alpha})^{-1}$. 
Note that the form of this value is similar to the limiting approximation.

\section{Numerical Results}
\begin{figure}[h]
\begin{center}
	\includegraphics[width=2.75in]{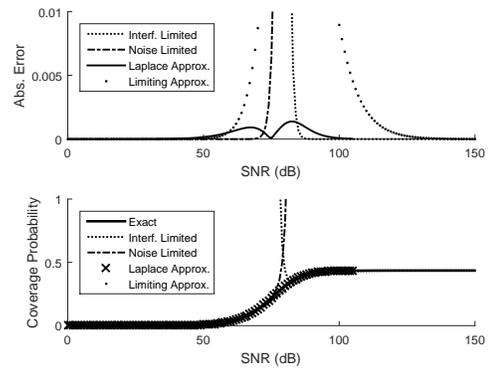}
	\caption{Error comparison when $\alpha = 3$. }
	\label{fig:alpha3}
 \end{center}
\end{figure}

\begin{figure}[h]
\begin{center}
	\includegraphics[width=2.75in]{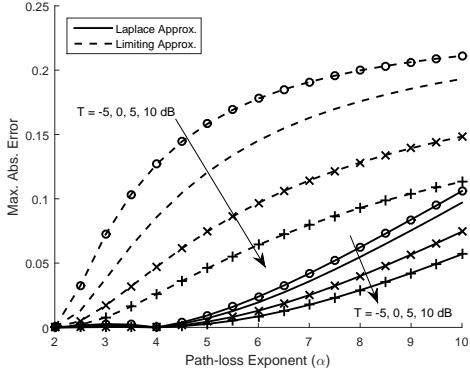}
	\caption{Maximum absolute error as $\alpha$ changes.}
	\label{fig:max_error_plot}
 \end{center}
\end{figure}

Here we compare the four approximations to integral (\ref{eqn:coverage}) with its value obtained via numerical integration. Since both parameters $A$ and $B$ depend on $\mu$ and $T$, we maintain them both at $\mu = T = 0$ dB. We assume $\lambda = 1/(\pi500^2)$, i.e., one BS on average per circular area of radius 500 meters. We only vary $\sigma^2$. The interferer's channel distribution $g$ is assumed to be exponential with mean $1/\mu$. Fig. 1 plots the change in absolute error and coverage probability as SNR = $1/\sigma^2$ changes for $\alpha=3$. For both the interference-limited and noise-limited cases, we take the summation up to four terms from (\ref{eqn:first-approx-series}) and (\ref{eqn:second-approx-series}), respectively. For the interference-limited case (\ref{eqn:second-approx-series}), the error is near zero only for SNR greater than certain value. Likewise, for the noise-limited case  (\ref{eqn:first-approx-series}), the error is near zero only for SNR smaller than some value. For intermediate case, the Laplace approximation gives an error less than 0.005.

Fig. 2 plots the maximum absolute error for the limiting and Laplace approximations. In general, the error amplitude tends to increase with $\alpha$, except for $\alpha=4$ where the error is zero for Laplace. The error amplitude decreases as $T$ increases, while it remain unchanged when $\lambda$ or $\mu$ is changed. This is because when $\lambda$ is increased, the system becomes interference limited at higher noise variance $\sigma^2$ (i.e., at lower SNR). Thus the coverage probability curve in Fig. \ref{fig:alpha3} shifts to the left. This shift is quantitatively given by (10) and (11). However, the maximum coverage probability, given by $1/\beta$, which occurs when $\sigma^2 \rightarrow 0$ (and thus $I \rightarrow 1/A$ at high SNR), remains unchanged. Similarly, since the transmit powers of all the BSs are assumed to be equal in \cite{Andrews2011},  changing the value of transmit power is equivalent to scaling the noise variance $\sigma^2$ by $\mu$ in the SINR expression. Thus, the plot of the coverage probability versus the $\sigma^2$ in dB results in shifting of the curves towards the left as $\mu$ is increased, without changing its shape or amplitude. Since the coverage probability curve has only shifted and not changed its shape, the maximum error which occurs around the edge of the plateau of the coverage probability will also remain unchanged.


\section{Conclusion}
We have examined four different ways of approximating the coverage integral. The limiting approximation is useful as an initial rough approximation. We can use the approximation for interference-limited case or the noise-limited case so long as the parameters $A$ and $B$ satisfy some inequality relationship. For intermediate cases, we recommend the use of Laplace approximation.


\appendices
\section{}
To check the convergence of the series (\ref{eqn:first-approx-series}) we will apply the ratio test: $\sum a_k$ converges if $\lim_{k\rightarrow\infty} |\frac{a_{k+1}}{a_k}|<1$.
 For our case,
 \begin{align*}
 \Big| \frac{a_{k+1}}{a_k} \Big| &= \frac{B}{(k+1)A^{\alpha/2}} \cdot \frac{\Gamma(\frac{(k+1)\alpha}{2}+1)}{\Gamma(\frac{k\alpha}{2}+1)}.
 \end{align*}
 Using the identity $B(x,y) = \frac{\Gamma(x)\Gamma(y)}{\Gamma(x+y)}$, we have
 \begin{align*}
 = & \frac{B \Gamma(\frac{\alpha}{2})}{A^{\alpha/2}} \cdot \frac{1}{k+1} \cdot \frac{1}{B(\frac{k\alpha}{2}+1,\frac{\alpha}{2})}.
\end{align*}
 Using the identity: $B(x+1,y)= B(x,y) \frac{x}{x+y}$, we obtain
 \begin{align*}
 = & \frac{B \Gamma(\frac{\alpha}{2})}{A^{\alpha/2}} \cdot \frac{1}{k} \cdot \frac{1}{B(\frac{k\alpha}{2},\frac{\alpha}{2})}.
 \end{align*}
For large $x$ and fixed $y$, $B(x,y) \sim \Gamma(y) x^{-y}$, so we have
\begin{align*}
\lim_{k\rightarrow\infty} \Big| \frac{a_{k+1}}{a_k} \Big| & = \frac{B \Gamma(\frac{\alpha}{2})}{A^{\alpha/2}} \cdot \lim_{k\rightarrow\infty} \frac{1}{k} \cdot \frac{1}{\Gamma(\frac{\alpha}{2})(\frac{k\alpha}{2})^{-\alpha/2}} \\
& = B \Big(\frac{\alpha}{2A}\Big)^{\alpha/2} \lim_{k\rightarrow\infty} \frac{k^{\alpha/2}}{k}. 
\end{align*}
Therefore, we finally have
 \[ \lim_{k\rightarrow\infty} \Big| \frac{a_{k+1}}{a_k} \Big| = B \Big(\frac{\alpha}{2A}\Big)^{\alpha/2} \lim_{k\rightarrow\infty} k^{\frac{\alpha}{2} - 1}.\]
 
 For $\alpha < 2$, $\lim_{k\rightarrow\infty} | \frac{a_{k+1}}{a_k} |  = 0$; hence, the series converges. For $\alpha = 2$, $\lim_{k\rightarrow\infty} | \frac{a_{k+1}}{a_k}| = \frac{B}{A}$. Therefore, the series converges if $\frac{B}{A} < 1$. For $\alpha > 2$, $\lim_{k\rightarrow\infty} | \frac{a_{k+1}}{a_k} |  = \infty$. Thus, the series always diverges.

\section{}
As in \textbf{Appendix A}, we can perform the ratio test for convergence of series (\ref{eqn:second-approx-series}). 
Using the identity $B(x,y) = \frac{\Gamma(x)\Gamma(y)}{\Gamma(x+y)}$, the ratio of consecutive terms are
\begin{align*}
\Big| \frac{a_{k+1}}{a_k} \Big| =& \frac{A \Gamma(\frac{4}{\alpha})}{(k+1)B^{2/\alpha} \Gamma(\frac{2}{\alpha})} \frac{B(\frac{2k}{\alpha},\frac{2}{\alpha})}{ B(\frac{2k}{\alpha}, \frac{4}{\alpha})}.
\end{align*}

For large $x$ and fixed $y$, $B(x,y) \sim \Gamma(y) x^{-y}$. Therefore,
\begin{align*}
\lim_{k \rightarrow\infty} \Big| \frac{a_{k+1}}{a_k} \Big| &=  \frac{A \Gamma(\frac{4}{\alpha})}{B^{2/\alpha} \Gamma(\frac{2}{\alpha})}  \lim_{k\rightarrow \infty} \frac{1}{(k+1)} \frac{\Gamma(2/\alpha)(2k/\alpha)^{-2/\alpha}}{\Gamma(4/\alpha)(2k/\alpha)^{-4/\alpha}} \\
&= \frac{A}{B^{2/\alpha}} \lim_{k \rightarrow \infty} \frac{1}{k+1} \frac{2k}{\alpha}^{-2/\alpha + 4/\alpha} \\
&= \Big(\frac{2}{\alpha}\Big)^{2/\alpha} \frac{A}{B^{2/\alpha}} \lim_{k \rightarrow \infty} \frac{k^{2/\alpha}}{k+1}. 
\end{align*} 

When $\alpha<2$, $\lim_{k \rightarrow\infty} | \frac{a_{k+1}}{a_k} | = \infty$; hence, the series diverges. When $\alpha=2$, $\lim_{k \rightarrow\infty} | \frac{a_{k+1}}{a_k} | = (\frac{2}{\alpha})^{2/\alpha} \frac{A}{B^{2/\alpha}}$; hence, the series converges if $(\frac{2}{\alpha})^{2/\alpha} \frac{A}{B^{2/\alpha}} < 1$. When $\alpha > 2$,  $\lim_{k \rightarrow\infty} | \frac{a_{k+1}}{a_k} | = 0$; hence, the series converges.

\bibliographystyle{IEEE}

\begin{thebibliography}{1}

\bibitem{Andrews2011}
J.G. Andrews, F. Baccelli, R.K. Ganti, ``A tractable approach to coverage and rate in cellular networks,'' {\em IEEE Trans. Commun.}, vol. 59,  no. 11, pp. 3122--3134, 13 October 2011.

\bibitem{Dhillon2012} 
H.S. Dhillon, R.K. Ganti, F. Baccelli, J.G. Andrews, ``Modeling and analysis of $K$-tier downlink heterogeneous cellular networks,'' {\em IEEE J. Sel. Areas Commun.}, vol. 30, no. 3, pp. 550--560, April 2012.

\bibitem{Goldsmith2005}
A. Goldsmith, {\em Wireless Communications}. Cambridge University Press, 2005.

\bibitem{Jeffrey-Zwillinger2007}
A. Jeffrey and D. Zwillinger, {\em Table of Integrals, Series, and Products
Table of Integrals, Series, and Products Series}. ed. 7, Academic Press, 2007.

\bibitem{Wong2001}
R. Wong, {\em Asymptotic Approximation of Integrals}. Society for Industrial and Applied Mathematics, 2001.

\end{thebibliography}

\end{document}